\documentstyle[prl,aps,epsf,multicol]{revtex}
\firstfigfalse
\begin{document}

\title{Concentration Dependence of the Effective Mass of $^3$He Atoms
in $^3$He-$^4$He Mixtures} 
\author{E.~Krotscheck,$^{\dagger\#}$
M.~Saarela$^\ddagger$, K. Sch\"orkhuber$^\dagger$, and R. Zillich$^\dagger$}
\address{$^\dagger$Institut f\"ur Theoretische Physik,
Johannes Kepler Universit\"at, A 4040 Linz, Austria}
\address{$^{\#}$Department of Physics,
Texas A\&M University, College Station, TX 77843, USA}
\address{$^\ddagger$Department of Physical Sciences, Theoretical
Physics, 
University of Oulu, SF-90570 Oulu, Finland}
\maketitle
\widetext
\begin{abstract}
Recent measurements by Yorozu {\it et al.\/} (S. Yorozu, H. Fukuyama,
and H. Ishimoto, Phys. Rev. B {\bf 48}, 9660 (1993)) as well as by
Simons and Mueller (R. Simons and R.~M. Mueller,
Czhechoslowak Journal of Physics Suppl. {\bf 46},
201 (1976))
have determined the effective
mass of $^3$He atoms in a $^3$He-$^4$He-mixture with great
accuracy. We here report theoretical calculations for the dependence
of that effective mass on the $^3$He concentration. Using correlated
basis functions perturbation theory to infinite order to compute
effective interactions in the appropriate channels, we obtain good
agreement between theory and experiment.
\end{abstract}
\begin{multicols}{2}
\narrowtext
The effective mass of a $^3$He atom in liquid $^4$He is due to two
effects. The dominant one is hydrodynamic backflow of the $^4$He
liquid around the impurity\cite{Lap,Feynman}.  This effect is largely
independent on the $^3$He concentration.  The second effect is the
dynamics imposed on the $^3$He component by the Pauli
principle\cite{BBP}. This effect causes a noticeable
concentration dependence which has recently been measured with high
accuracy by Yorozu {\it et al.\/}\cite{Yorozu93} and by Simons
{\it et al.\/}\cite{SimonsLT21}; it is the subject of this paper.

Ground state properties of $^3$He- $^4$He mixtures like the energetics
of the system and its local structure are today quite well understood
from a microscopic point of view\cite{MixMonster}.  For a microscopic
understanding of the {\it hydrodynamic\/} effective mass, rather
advanced methods are needed\cite{SKJLTP} due to the high density of
the $^4$He background.  On the other hand, the $^3$He component in the
mixture is dilute, and the interaction between individual $^3$He atoms
is dominated by phonon exchange. Therefore, much simpler methods than
those necessary\cite{KSCP81} for obtaining reliable results in pure $^3$He
are adequate for calculating ground state properties of dilute
mixtures\cite{MixMonster}.

Microscopic many-body theory postulates an empirical
Hamiltonian
\begin{equation}
H = -\sum_\alpha\sum_{i=1}^{N_\alpha}
{\hbar^2\over 2m_\alpha}\nabla_i^2 +{1\over 2}\sum_{\alpha\beta}
\mathop{{\sum}'}_{i,j}^{N_\alpha, N_\beta}
V(|{\bf r}_i^{(\alpha)}- {\bf r}_j^{(\beta)}|)
\label{hamiltonian}
\end{equation}
that contains only a  two-body interaction\cite{Aziz}
and the masses of the two types of particles.
One then uses the Feenberg form\cite{FeenbergBook,Chuckreview}
for the ground state wave function,
\begin{eqnarray}
\Psi_0(\{{\bf r}_i^{(\alpha)}\})
	&&=e^{{1\over 2} U(\{{\bf r}_i^{(\alpha)}\})}
	\Phi_0(\{{\bf r}_i^{(3)}\}),\nonumber\\
	U(\{{\bf r}_i^{(\alpha)}\})&& = {1\over 2!}\sum_{\alpha\beta}
	\mathop{{\sum}'}_{i,j}^{N_\alpha, N_\beta}
	u^{\alpha\beta}({\bf r}_i,{\bf r}_j)\nonumber\\
	&&+{1\over 3!}\sum_{\alpha\beta\gamma}
	\mathop{{\sum}'}_{i,j,k}^{N_\alpha,N_\beta,N_\gamma}
	u^{\alpha\beta\gamma}
	({\bf r}_i,{\bf r}_j,{\bf r}_k),
\label{groundstate}
\end{eqnarray}
where $\Phi_0(\{{\bf r}_i^{(3)}\})$ is a Slater determinant of plane
waves ensuring the antisymmetry of the Fermion component of the
mixture. The superscripts $\alpha$, $\beta$, $\ldots$  refer to the
type of correlated particles; the prime on the summation symbol in
Eqs.~(\ref{hamiltonian}), (\ref{groundstate}) indicates that no
two pairs $(i,\alpha)$, $(j,\beta)$ can be the same. The
correlation functions $u^{\alpha\beta}({\bf r}_i,{\bf r}_j)$ and
$u^{\alpha\beta\gamma}({\bf r}_i,{\bf r}_j,{\bf r}_k)$ are
determined by the functional minimization of the ground state
energy
\begin{equation}
	E_0 = \langle\Psi_0 | H | \Psi_0\rangle/
	\langle\Psi_0 | \Psi_0\rangle
\end{equation}
with respect to the correlation functions\cite{PPA1,PPA2,EKthree}.
Details of the procedure have been described in
Ref. \onlinecite{MixMonster}.

%
Within the above variational theory, the single particle
excitation spectrum is calculated by allowing for an occupation of
single particle orbitals $n_k$ in the Slater function that is
different from the Fermion ground state $n_k^{(0)} = \theta(k_F-k)$,
where $k_F$ is the Fermi momentum of the $^3$He component. The
single particle spectrum can then be calculated by the variation
\begin{equation}
\epsilon^{(3)}(k)
 = {\delta E_0 \over \delta n_k} = t_3 (k) + u(k) + U_0\ ,
\label{epsHF}
\end{equation}
where we use $t_\alpha(k) = \hbar^2 k^2/2m_\alpha$ for the kinetic
energy of a
free particle of species $\alpha$. $U_0$ is a constant related to the
chemical potential, and $u(k)$ a momentum dependent average field. The
general graphical analysis of the variational single-particle field
$u(k)$ has been carried out in Ref. \onlinecite{CBF2}; if the $^3$He
component is dilute, the average field can be written in the form of a
Hartree-Fock field in terms of a local, effective interaction
$\tilde W_{\rm eff}(q)$
\begin{equation}
u(k) = - \int {d^3 q\over (2\pi)^3\rho_3}n_q^{(0)}
\tilde W_{\rm eff}(|{\bf k-\bf q}|)\ .
\label{pFHNC}
\end{equation}


It is known\cite{KSCP81} that the na\"\i ve use of a variational wave
function of the type (\ref{groundstate}) leads to an effective mass of
{\it pure\/} $^3$He that is --- in sharp contrast to
experiments\cite{Wheatley75,AHMS80} --- less than
one\cite{IndianMass}.  The cause of this deficiency is that the wave
function (\ref{groundstate})) describes the {\it average\/}
correlations between particles, but it is not sensitive to the {\it
specifics\/} of the correlations in the vicinity of the Fermi surface.
The cure for the problem is Correlated-Basis Functions (CBF)
theory\cite{FeenbergBook} to infinite
order\cite{Bengt}. The theory can be mapped on a Green's
function approach in terms of effective interactions\cite{rings} that
are provided by the variational theory.  The single particle
properties are described by a complex {\it self-energy\/}
$\Sigma(k,\omega)$; the single particle spectrum $\varepsilon(k)$ is
obtained from the solution of equation
\begin{equation}
\varepsilon(k) = t_3(k) + \Sigma(k,\varepsilon(k))\ .
\label{selfen}
\end{equation}
If only one-phonon coupling processes are considered, the self-energy
$\Sigma(k,E)$ is given by the so-called
G0W-approximation\cite{Hedin65,Rice65}
\begin{equation}
\Sigma(k,E) = i \int { d^3 q d(\hbar\omega)\over (2\pi)^4\rho_3}
G^{(0)}(\left|{\bf k}-{\bf q}\right|,{E\over\hbar}-\omega)
\tilde V_{\rm eff\/}(q,\omega).
\label{G0W}
\end{equation}
\begin{equation}
G^{(0)}(k,\omega) = {1-n_k^{(0)}\over \hbar\omega - t_3(k) + i\eta} 
+ {n_k^{(0)}\over \hbar\omega - t_3(k) - i\eta} 
\label{green}
\end{equation}
is the free single-particle Green's function and
\begin{equation}
\tilde V_{\rm eff\/}(q,\omega) = 
\tilde V_{\rm p-h}^{33}(q) + \sum_{\alpha\beta}
\tilde V_{\rm p-h}^{3\alpha}(q)\chi_{\alpha\beta}(q,\omega)
\tilde V_{\rm p-h}^{3\beta}(q)
\label{VeffE}
\end{equation}
is the effective, {\it energy dependent\/} $^3$He-$^3$He interaction.
In Eq. (\ref{VeffE}), $\tilde V_{\rm p-h}^{\alpha\beta}(q)$ is
the local, {\it particle-hole irreducible\/}
interaction matrix\cite{MixMonster}, and $\chi_{\alpha\beta}(q,\omega)$ is
the density-density response matrix. The particle-hole irreducible
interactions, which are in conventional Green's functions theories the
most significant source of uncertainty, are provided by the
variational ground state theory.

To separate the ``hydrodynamic'' and the ``fermionic'' component
of the self-energy, we rewrite the single-particle Green's function
as
\begin{eqnarray}
G^{(0)}(k,\omega) &=& {1\over \hbar\omega - t_3(k) + i\eta}
\nonumber\\ 
&+& n_k^{(0)}\left[{1\over  t_3(k) - \hbar\omega - i\eta} 
- {1\over t_3(k) -\hbar\omega + i\eta}\right]\nonumber\\
&\equiv& G_H^{(0)}(k,\omega) + G_F^{(0)}(k,\omega)
\label{green2}
\end{eqnarray}
and, correspondingly, the self-energy as
\begin{equation}
\Sigma(k,E) = \Sigma_H(k,E) + \Sigma_F(k,E)\ .
\label{SigmaF}
\end{equation}
The ``hydrodynamic'' part $\Sigma_H(k,E)$ of the self-energy is, in the
limit $\rho_3\rightarrow 0$,
identical to the self-energy of a single $^3$He impurity in the $^4$He
host liquid ({\it cf.\/} Eq. (3.30) of Ref. \onlinecite{SKJLTP});
since we have found in Ref. \onlinecite{MixMonster} that the
concentration dependence is generally weak it is appropriate to
identify $\Sigma_H(k,E)$ with the self-energy of a single $^3$He atom
also at finite concentrations; we will return to this quantity later.


Let us first focus on the {\it second\/} term in Eq. (\ref{SigmaF}).
The energy integration yields the compact form
\begin{equation}
\Sigma_F(k,E) = -\int {d^3 q\over (2\pi)^3\rho_3}
n_q^{(0)}\tilde V_{\rm eff\/}({\bf k}-{\bf q},E-t_3(q))\ .
\label{SigmaFinal}
\end{equation}
The relationship to Eq. (\ref{pFHNC}) is apparent: If $\tilde V_{\rm
eff\/}(q,\omega)$ were energy independent, $\Sigma_F(k,\omega)$ would
be just a Hartree-Fock average field of the form
(\ref{pFHNC}). Indeed, the expression (\ref{pFHNC}) can be derived
from Eq.  (\ref{SigmaFinal}) using the same ``average-energy''
procedure that has been employed to establish the connection between
the parquet-diagram theory and the optimized HNC
theory\cite{parquet2}, namely to identify
\begin{equation}
\tilde W_{\rm eff}(q) = \tilde V_{\rm eff\/}(q,\bar\omega(q))
\label{equiv}
\end{equation}
where the average energy $\hbar\bar\omega(q)$ is chosen such that both the
energy dependent and the energy independent interactions produce the
same static structure function. We stress that this result
is only an {\it observation\/} on how the static approximation
and the G0W approximation are related. It does
{\it not\/} imply that this approximation is also {\it
adequate\/} for the single-particle properties.


With the relationship between theories established, we now turn to the
numerical application. The ingredients of the theory --- the effective
interactions $\tilde V_{\rm p-h}^{\alpha\beta}(q)$ and $\tilde W_{\rm
eff}(q)$ as well as the Feynman spectrum $\epsilon^{(4)}(k)$ have been
obtained in Ref. \onlinecite{MixMonster}. The first important quantity
is the {\it hydrodynamic effective mass.\/} Due to the high density of
the background more elaborate methods than the G0W approximation must
be used for a quantitative prediction\cite{SKJLTP}. To identify the
Fermi-liquid effects we are interested in here, we have, however, not
used these results, but rather let the hydrodynamic mass be a free
parameter. After the concentration dependence was calculated from the
Fermi-liquid contributions we made a single parameter fit to the
experiments of Refs. \onlinecite{Yorozu93} and \onlinecite{SimonsLT21}
to optimize the {\it overall\/} agreement. That way, we arrived at the
following interpolation formulas for the hydrodynamic mass:
\begin{equation}
        \left.{m_H\over m_3}\right)_{\rm expt} = 
	2.171 + 2.334 r + 4.547 r^2  - 9.329 r^3
\label{japmass}
\end{equation}
for the data of Ref. \onlinecite{Yorozu93} and
\begin{equation}
        \left.{m_H\over m_3}\right)_{\rm expt} = 
	2.124 + 2.630 r + 2.000 r^2
\label{julmass}
\end{equation}
from those of Ref. \onlinecite{SimonsLT21}. Here,
$r = \rho_4/\rho_0-1$, $\rho_4$ is the
$^4$He density and $\rho_0$=0.02183\AA$^{-3}$ is its value at the
saturation vapor pressure. Typically, the discrepancy between the two
different extrapolations is 0.03, these values are throughout the full
density regime about 0.1 above our theoretical calculation of
Ref. \onlinecite{SKJLTP}.

A possible complication to be considered is the momentum dependence of
the hydrodynamic effective mass since the particles at the Fermi
surface have a finite momentum. For that purpose, we have calculated the
single-impurity spectrum,
\begin{equation}
	\hbar\omega(k) = t_3(k) + \Sigma_H(k,\omega(k))\ ,
\end{equation}
and have determined the momentum dependence of the
hydrodynamic mass by writing, in a momentum regime $0 ~\le k\le
0.4~{\rm\AA}^{-1}$ (chosen such that the upper boundary
is comparable to the Fermi momentum at the highest pressure and
concentration) the spectrum in the form
\begin{equation}
	\hbar\omega_H(k) = {\hbar^2 k^2\over 2m_H} (1+b k^2)\ .
\end{equation}
We have verified that this form is adequate at small momenta and has a
weak density dependence in
agreement with experiments. \cite{Fak,Owers88} We have used the value
$b=-0.074$ at all densities. 

\begin{figure}
\noindent\epsfxsize=2.6truein\epsffile{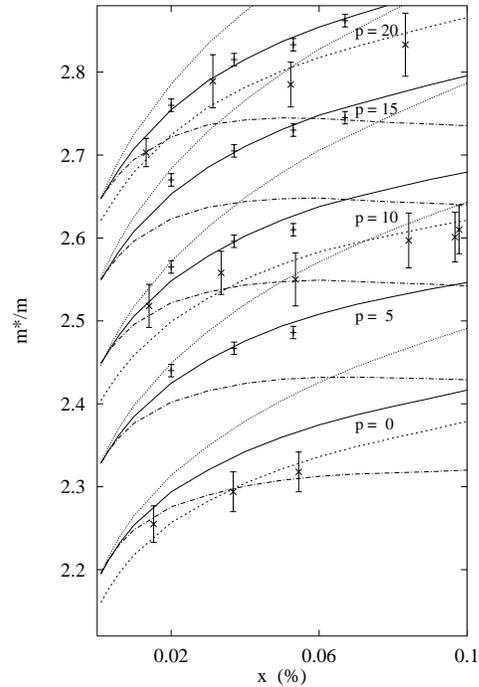}
\caption{Theoretical and experimental effective mass ratio
$m^*(P,x)/m$ as a function of pressure $P$ and concentration
$x$. The thick full curve is the fully self-consistent result
with the hydrodynamics mass fitted to the experiments of
Ref. \protect\onlinecite{Yorozu93} (circles with error bars). The thin,
dashed-dotted curve is the on-shell approximation and the thin, short dashed
curve the static approximation using the same hydrodynamic  mass. The
thick dashed curve is the fully self-consistent result using the
hydrodynamic  mass fitted to the results of Ref. 
\protect\onlinecite{SimonsLT21} (boxes with error bars).
\label{massxj}}
\end{figure}

Three calculations have been carried out to determine the Fermi-liquid
contributions to the effective mass of the $^3$He component as a
function of concentration
and density. The transformation from the density and concentration
dependence to the pressure and concentration is done using
experimental results given in Ref.~ \onlinecite{OuY}.
The first is the simple
approximation (\ref{pFHNC}). To account for the hydrodynamic
backflow, one must supplement the Fermion contribution (\ref{epsHF})
by the hydrodynamic contribution; the spectrum has the form
\begin{equation}
\epsilon^{(3)}(k) = \hbar\omega_H(k) + u(k) + U_0
\label{epsHNC}
\end{equation}
where the Fermi correction $u(k)$ is given in Eq.
(\ref{pFHNC}). The effective masses derived from this
spectrum are significantly above the experimental one, {\it cf.\/}
Fig.~\ref{massxj}.

In the next step, we use the full self-energy in an ``on-shell''
approximation
\begin{equation}
\epsilon_{\rm OS}^{(3)}(k) = \hbar\omega_H(k) + \Sigma_F(k,t_3(k))\ .
\label{epsONSH}
\end{equation}
This form of the self-energy relaxes the approximations made by the
variational theory, we see in Fig.~\ref{massxj} that the agreement
with the experiment is no better than in the FHNC approximation;
the on-shell approximation, the effective mass now being significantly
{\it below\/} the experimental value.

Finally, we carry out a self-consistent calculation of the effective
mass. Due to the low concentrations, we may assume a single-particle
spectrum of the form $t_3(k) = \hbar\omega_H(k)$ in the Green's
function (\ref{green}) and, consequently, in Eq. (\ref{SigmaFinal});
note that the hydrodynamic mass is included in the Green's
function. This effective mass is then determined self-consistently by
requiring that the spectrum $\epsilon^{(3)}(k)$ determined by
\begin{equation}
\epsilon^{(3)}(k) = \hbar\omega_H(k) +
\Sigma_F\left(k,\hbar\omega_H(k)\right)
\label{epsFULL}
\end{equation}
can be fitted by the same effective mass that has been used in the
self-energy. This theoretical calculation produces, especially at
lower densities, a slightly stronger concentration dependence than
seen experimentally. Compared with both the  FHNC and the
``on-shell'' approximation, the self-consistent result shown in
Fig.~\ref{massxj} appear quite satisfactory.

The essential difference between the on-shell and the self-consistent
calculation is that information about the hydrodynamic backflow has
gone into $G_F^{(0)}(k,\omega)$. Note that hydrodynamic backflow is
--- as a feature of excitations --- generically {\it not\/} included
in the approximation (\ref{pFHNC}). Indeed, generalizing the
variational theory to dynamic correlations {\it without\/} introducing
the coupling to the background would lead to an effective mass close
to the on-shell mass, which is notably too low.

To produce Fig. \ref{massxj} we have used -- as stated
before -- the hydrodynamic mass given in Eqs. (\ref{japmass}) or
(\ref{julmass}), respectively. Our calculations predict a curvature of
the effective mass as a function of concentration, leading to a
hydrodynamic effective mass slightly lower than the one predicted by
linear extrapolation, {\it cf.\/} Table \ref{hydmass}. Such a
curvature is caused by the Fermi-functions, already the simple
approximation (\ref{epsHNC}) predicts a behavior
\begin{equation}
	m^*(x) = m_H + a x^{2/3} + b x + c x^{5/3} +d x^{7/3}\ .
\label{massfit}
\end{equation}
The precise values of these coefficients is a matter of a microscopic
calculation.  In Table \ref{hydmass} we list the values of $a$, $b$,
and $c$ for different pressures as obtained from the least square fit
to the self-consistent solution of Eq. (\ref{epsFULL}).


We have demonstrated that the concentration dependence of the
effective mass is essentially a Fermi-liquid effect, enhanced by the
renormalization of the single particle propagator through hydrodynamic
backflow. The technical simplifications caused by the relatively low
density of the Fermion component has allowed to highlight the relevant
physical mechanisms quite clearly. The data of Ref.
\onlinecite{SimonsLT21} are consistently below those of Ref.
\onlinecite{Yorozu93}; the difference of 0.03 might be attributable to the
different pressure gauge. This difference might appear negligible, but
we need to point out that it produces uncomfortably large
uncertainties in predictions for the first antisymmetric Landau
parameter $F_0^a$ from magnetic susceptibility measurements\cite{APCT76}.

On the other hand, we find that the concentration dependence of the
effective mass --- in other words the generic Fermi-liquid effect
--- is consistent between both sets of experiments and theory.
Extrapolations to zero concentration also appear to be consistent.

Besides providing an accurate microscopic calculation of the
concentration dependence of the $^3$He effective mass in $^4$He, we
have analyzed various procedures for calculating the effective mass.
By comparing the static calculation with the dynamic CBF calculation,
we have discussed an instructive example for the delicate interplay
between single-particle and Fermi-liquid effects in $^3$He-$^4$He
mixtures.

The work was supported, in part, by the National Science Foundation
under grant DMR-9509743, the Austrian Scinece Fund
under project P11098-PHY, and the
Academy of Finland. Discussions with and numerous comments by
C. E. Campbell as well as useful comments and communication of
unpublished data by R. M. Mueller and R. Simons are gratefully
acknowledged.

\end{multicols}
\widetext
\begin{table}

\caption{Pressure dependence of the
hydrodynamic effective mass and the coefficients of the expansion
(\protect\ref{massfit}) as obtained from the present
calculation and fitted to the data of 
Ref. \protect\onlinecite{Yorozu93}. Also shown are
the linear extrapolation of Ref. \protect\onlinecite{SimonsLT21}
as well as our fit to these data.}
\begin{tabular}{ccccccccc}
$P$~~(Atm)  & $m_H$ (this work) & $a$ & $b$ & $c$ & $d$ & $m_H$
Ref. \protect\onlinecite{SimonsLT21} & $m_H$ fitted to
Ref. \protect\onlinecite{SimonsLT21} data \\
\hline
0  &  2.18 &  1.49 & 1.39 & - 18.2 & 36.7 & 2.23$\pm$0.02 & 2.15 \\
5  &  2.31 &  1.07 & 3.00 & - 22.6 & 40.2 && \\
10 &  2.44 & 0.789 & 4.48 & - 28.2 & 50.4 & 2.52$\pm$0.02 & 2.39 \\
15 &  2.54 & 0.501 & 6.17 & - 36.1 & 66.8 &&\\
20 &  2.64 & 0.310 & 7.41 & - 42.1 &80.1& 2.70$\pm$0.03 & 2.62 \\
\end{tabular}
\label{hydmass}
\end{table}
\end{document}